# Dual Optical Hyperbolicity of PdCoO$_2$ and PdCrO$_2$ Delafossite Single Crystals

*Salvatore Macis\*, Annalisa D'Arco, Eugenio Del Re, Lorenzo Mosesso, Maria Chiara Paolozzi, Vincenzo Stagno, Alexander McLeod, Yu Tao, Pahuni Jain, Yi Zhang, Fred Tutt, Marco Centini, Maria Cristina Larciprete, Chris Leighton, Stefano Lupi*

S. Macis, A. D'Arco, E. Del Re, L. Mosesso, M.C. Paolozzi, S. Lupi
Department of Physics, Sapienza University, Piazzale Aldo Moro 5, 00185 Rome, Italy
E-mail: salvatore.macis@uniroma1.it

V. Stagno
Department of Earth Sciences, Sapienza University, Piazzale Aldo Moro 5, 00185 Rome, Italy

A. McLeod
School of Physics and Astronomy, University of Minnesota, 421 Washington Avenue SE, Minneapolis, MN 55455, United States

M. Centini, M. C. Larciprete
Dipartimento di Scienze di Base e Applicate per l'Ingegneria, Sapienza Università di Roma, Via A. Scarpa 16, 00161 Roma, Italy

Y. Tao, P. Jain, Y. Zhang, F. Tutt, C. Leighton
Department of Chemical Engineering and Materials Science, University of Minnesota, 421 Washington Avenue SE, Minneapolis, MN 55455, United States

**Abstract:**

Hyperbolic materials exhibit a very peculiar optical anisotropy with simultaneously different signs of the dielectric tensor components. This anisotropy allows the propagation of exotic surface-wave excitations like hyperbolic phonons and plasmon polaritons. While hyperbolic materials hold promise for applications in subwavelength photonics and enhanced light-matter interactions, their natural occurrence is limited to few materials, often accompanied by significant dielectric losses and limited hyperbolic spectral bandwidth. Focusing on PdCoO$_2$ and PdCrO$_2$ delafossite transition-metal oxides, in this paper we demonstrate their unique dual hyperbolic regimes: one localized around a phonon absorption in the mid-infrared spectral region, and the other extending into the visible range. Both hyperbolic regimes show exceptional properties including low dissipation and high hyperbolic quality factors. These results pave the way for innovative applications of delafossite layered metals in subwavelength photonics, imaging, and sensing.

## 1. Introduction

The electrodynamics of a hyperbolic optical system are characterized by an anisotropic dielectric tensor, whose diagonal components $\epsilon_{xx}$, $\epsilon_{yy}$, $\epsilon_{zz}$ have opposite signs within a given spectral range. Specifically, Type I hyperbolic materials have one component of the dielectric tensor which is negative[1–3], while Type II hyperbolic materials have two negative components (the so called indefinite-material). As a result, hyperbolic materials can support the propagation of high-wavevector electromagnetic surface-waves through the excitation of surface plasmon polariton (SPP) or surface phonon polariton (SPhP) modes[4–6]. In recent years, hyperbolic optical materials have gained significant attention due to these unique electrodynamics properties, which could enable applications in subwavelength photonics, super-resolution imaging, spontaneous emission control, negative refraction, and enhanced light-matter interactions[1–3].

Hyperbolic optical properties are observed, however, only in a limited number of materials. One can cite, for instance, hexagonal boron nitride (hBN), graphite, cuprate oxides, magnesium diboride, topological insulators, sapphire, and other exotic materials[1–5,7–9]. Those systems are typically characterized by a narrow spectral range in which hyperbolicity manifests, accompanied by large imaginary parts of the dielectric tensor in the same frequency interval, resulting in strong losses. These limitations reduce the possible applications of these materials. In order to extend and control hyperbolic properties, research has recently been focused on the development of hyperbolic metamaterials[1–3]. Hyperbolic metamaterials have been engineered at the micro and nano spatial scales, using artificial structures often based on transparent conductive oxides [1–3,10]. Nano or micro artificial structuring enables precise control over dielectric properties, reducing losses and extending the Hyperbolic Operational Spectral Bandwidth (HOSB). However, a very broad HOSB with low losses, extended towards the technologically important infrared (IR) spectral range, remains a key challenge for applications of hyperbolic systems.

$PdCoO_2$ and $PdCrO_2$ transition-metal oxides belonging to the delafossite family[11–16] are characterized by a layered crystal structure with triangularly coordinated Pd layers, separated by $CoO_2$ ($CrO_2$) octahedra (see the inset in Figure 1c), exhibiting a remarkable 2-Dimensional (2D) electronic character. As a consequence, $PdCoO_2$ and $PdCrO_2$ display a similar room-temperature *ab* plane DC electrical conductivity, larger than alkali metals and as high as those of copper and silver, while showing lower DC conductivity perpendicular to the plane, i.e., along the *c* axis[15–17]. $PdCoO_2$ shows a paramagnetic character down to low temperature [15], while $PdCrO_2$ presents an antiferromagnetic transition around 38 K, remaining metallic at lower temperature[17].

In this paper, we measure the dielectric tensors of $PdCoO_2$ and $PdCrO_2$ at room temperature. We show that both materials present excellent Type II hyperbolic optical properties in a very broad spectral range from infrared to visible, additionally characterized by very low dissipation and high hyperbolic quality factors. Through the experimental dielectric tensors, we calculated both the hyperbolic frequency/wavevector relation and the hyperbolic plasmon- and phonon-polariton dispersion. Due to their strong anisotropy and their natural hyperbolicity, the investigated materials offer rich physical phenomena enabling the exploration of novel optical properties and unpredicted functionalities. This opens the way to innovative applications of delafossite layered oxides in nanophotonics, imaging, and sensing.

## 2. Results and Discussion

$PdCoO_2$ and $PdCrO_2$ single crystals studied in this paper were grown by chemical vapor transport (CVT)[11,12]. First, precursor powders were prepared using the metathesis/flux method [13,14,16], then the resulting phase-pure $PdCoO_2$ and $PdCrO_2$ precursor powders were loaded into evacuated quartz vessels for CVT crystal growth in a two-zone tube furnace, utilizing $PdCl_2$ as the transport agent (see Supporting Information for full details). Single crystals of several mm size were obtained, and cleaned of excess chlorides[11,12]. Energy-dispersive X-ray spectroscopy (EDS) and X-ray diffraction (XRD) were employed to assess the crystal quality and extract their lattice parameters, which were in agreement with literature[18,19] (see Figure S1 in Supporting Information). The two single crystals of $PdCoO_2$ and $PdCrO_2$ used in this paper have a size of ~200 μm along the *c* axis, and an area of ~1 $mm^2$ in the *ab* plane. In order to perform specular reflectance measurements, all crystals were polished with diamond powder to a surface roughness less than 300 nm.

The reflectance $R(\omega)$ of both $PdCoO_2$ and $PdCrO_2$ crystals was measured (see Figure 1) at room temperature in a wide spectral range from terahertz (THz) to visible (VIS) using a Bruker Vertex 70V Fourier-transform interferometer coupled to a Hyperion 1000 microscope (see Methods). Measurements were performed on the crystal surface parallel to the *ab* plane and along the *c* axis. We verified that the optical response perpendicular to the *c* axis (onto the surface containing the same axis) is the same as the one measured directly on the *ab* plane. The optical response along the *c* axis (panels a and b in Figure 1, corresponding to $PdCoO_2$ and $PdCrO_2$ respectively) and in the *ab* plane (panels c and d respectively), are extremely different. Both materials show a very high metallic reflectance in the *ab* plane with a plasma edge located in the near-IR, and a much more insulating response along the *c* axis corresponding to a small Drude term superimposed on strong phonon absorption below 1000 cm$^{-1}$. The in-plane $PdCoO_2$ reflectance measured in this work is in very good agreement with that of Homes *et al.* [15].

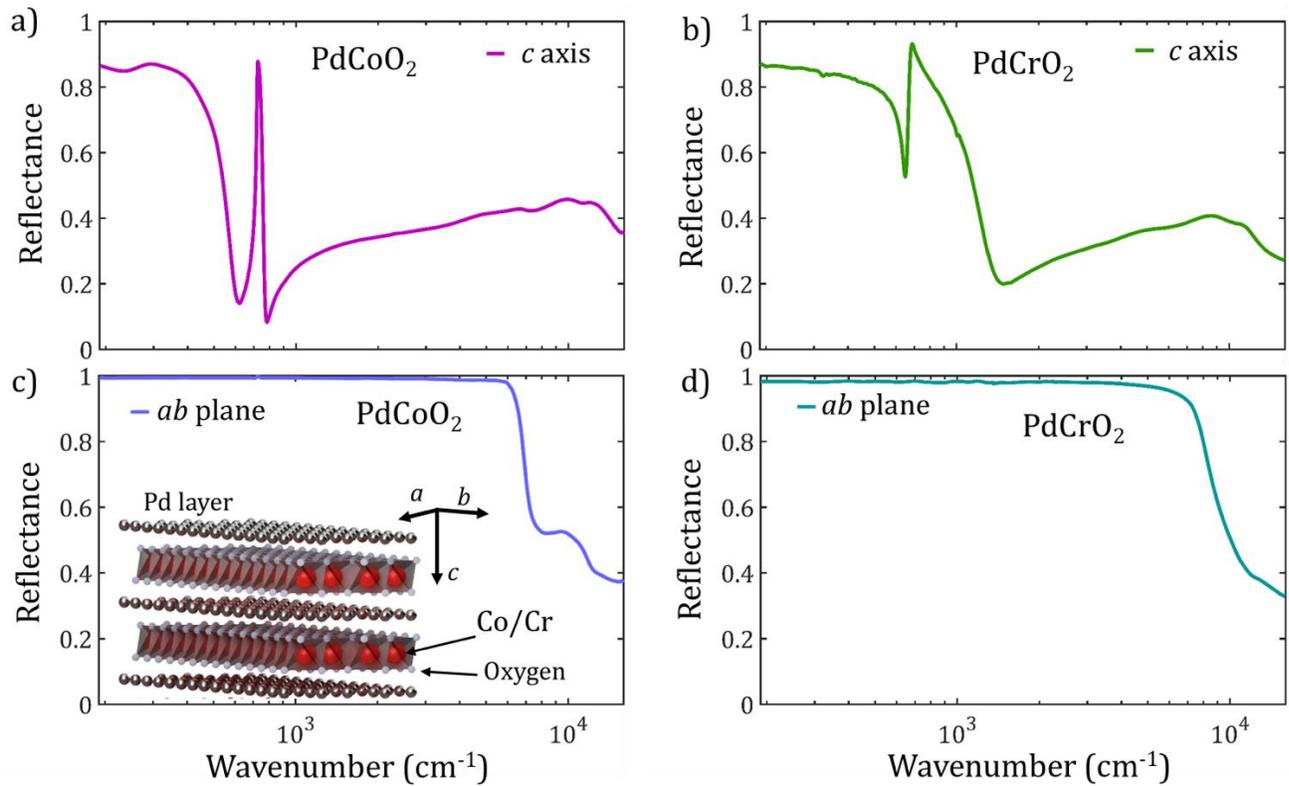

**Figure 1.** Reflectance spectra of $PdCoO_2$ (a,c) and $PdCrO_2$ (b,d) single crystals, measured with polarization along the *c* axis (a,b) and in the *ab* plane (c,d). The inset in panel c) schematically shows the layered structure of both materials, illustrating the Pd triangularly coordinated layers and $CoO_2$ ($CrO_2$) octahedra. The *c* axis reflectance presents a small Drude contribution superimposed on strong phonon signatures below 1000 cm$^{-1}$. For polarization parallel to the *ab* planes, one observes instead a strong metallic contribution with a plasma edge in the near-IR at around 10000 cm$^{-1}$.

The analysis of this strong optical anisotropy is the fundamental key for assessing and quantifying the hyperbolic nature of the investigated crystals. Reflectance spectra were analyzed with the RefFIT software, which employs a Drude-Lorentz Kramers-Kronig-constrained modeling of the material optical response[20]. This procedure provides all microscopic optical response functions. In particular, in Figure 2 we show the real and imaginary part of the dielectric function for both materials and polarizations.

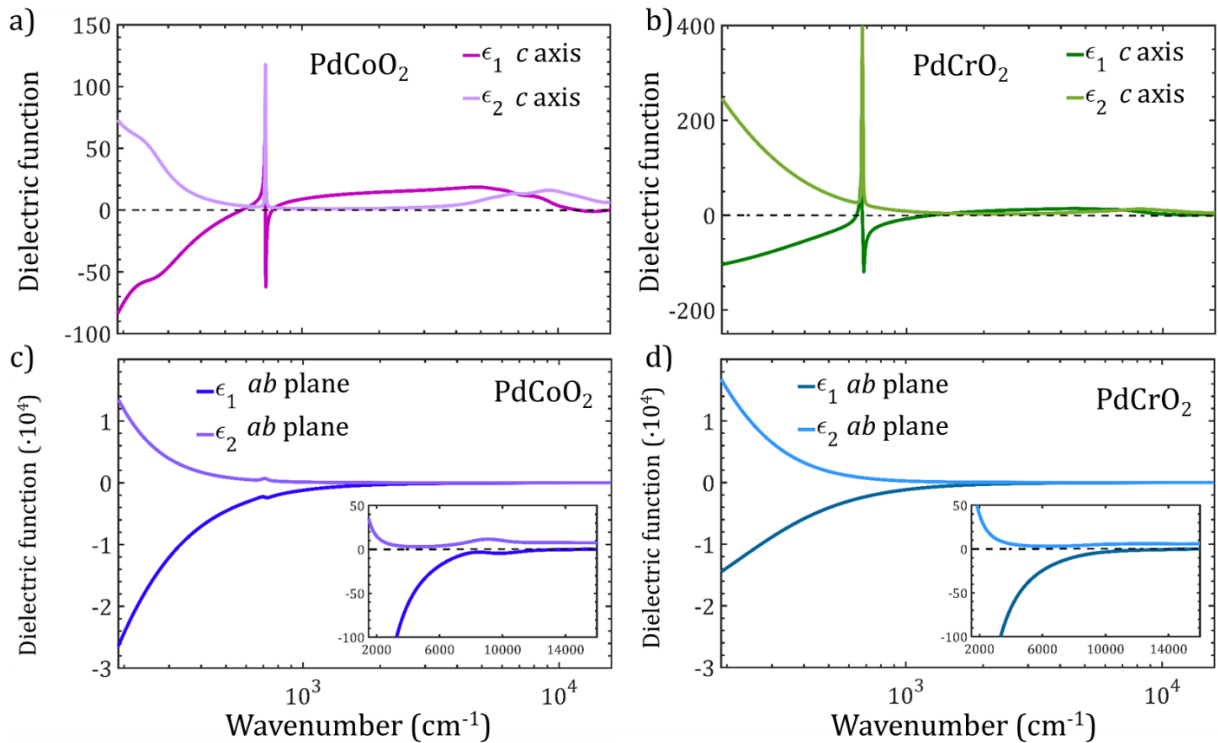

**Figure 2.** Complex dielectric functions of $PdCoO_2$ (a,c) and $PdCrO_2$ (b,d) extracted from the RefFIT fitting procedure (see text). Panels a) ($PdCoO_2$) and b) ($PdCrO_2$) show the real $\varepsilon_1$ and imaginary $\varepsilon_2$ part of the dielectric function with polarization parallel to the *c* axis. Here one observes a small metallic contribution in the THz range that can be described by a Drude term (see Table 1) and two peaks related to $A_u$ phonons[15,21]. In panels c) ($PdCoO_2$) and d) ($PdCrO_2$), for polarization parallel to the *ab* plane, one observes a strong Drude contribution related to the high DC conductivity of these materials. The two insets to panels c) and d) show a magnified view of the dielectric functions in the IR-VIS spectral region.

Table 1 reports the Drude parameters (plasma frequency $\omega_p$ and scattering rate $\Gamma$) as obtained from the fitting process. For both crystals, the plasma frequency increases by an order of magnitude from the $c$ axis to the $ab$ plane. The DC conductivity $\sigma_0$ extrapolated from the optical values of $\omega_p$ and $\Gamma$ obtained from this fit, are compatible with the values in literature estimated by resistivity data, with an anisotropic conductivity ratio $\sigma_{0,ab}/\sigma_{0,c}$ of ~230 and ~100 for PdCoO$_2$ and PdCrO$_2$, respectively[11,12,15,17,21,22]. Moreover, as shown in Tables S1 and S2 of Supporting Information, the dielectric tensors of both PdCoO$_2$ and PdCrO$_2$ are modulated by two $A_u$ phonons along the $c$ axis as predicted by group theory[15,21], as well as by some interband electronic transitions in the near-IR/VIS range. Their frequencies (see SI) are in good agreement with calculations[15,21,23].

| PdCoO$_2$ | | | |
|---|---|---|---|
| | $\omega_p$ (cm$^{-1}$) | $\Gamma$ (cm$^{-1}$) | $\sigma_0$ ($\Omega^{-1}$cm$^{-1}$) |
| $c$ axis | 2300 | 98 | 910 |
| $ab$ plane | 34 700 | 97 | 206 850 |
| PdCrO$_2$ | | | |
| | $\omega_p$ (cm$^{-1}$) | $\Gamma$ (cm$^{-1}$) | $\sigma_0$ ($\Omega^{-1}$cm$^{-1}$) |
| $c$ axis | 4240 | 390 | 970 |
| $ab$ plane | 35 010 | 220 | 92 900 |

**Table 1.** Drude ($\omega_p$ and $\Gamma$) parameters and optically extrapolated DC conductivity $\sigma_0$ of PdCoO$_2$ and PdCrO$_2$ extracted from the fitting of the reflectance spectra in Figure 1, for light polarization parallel to the $c$ axis and in the $ab$ plane.

Optical hyperbolicity in a material depends on the electrodynamic mechanisms generating the anisotropy and the sign of the dielectric tensor components. A Drude-Lorentz hyperbolicity, is given by metallic and insulating behavior, resulting in a Drude dispersion and a Lorentz one along different crystalline axes. Similarly, we can define a Drude-Drude hyperbolicity when we have two different Drude contributions along different axes, and a Lorentz-Lorentz type when the electrodynamic response is dielectric along all crystalline orientations[24,25]. The latter ones are also known as phononic hyperbolic materials.

Both delafossite PdCoO$_2$ and PdCrO$_2$ metals present (see above discussion and Figures 1 and 2) a strong 2D electrodynamic behavior in which $\epsilon_{1,ab}$ is characterized by a highly metallic negative value, while $\epsilon_{1,c}$, in the same spectral interval, has two separate positive regions. The first one is related to the strong phononic resonance below 1000 cm$^{-1}$. After this phonon region, the small Drude contribution of both materials phases out, and another sign crossing of $\epsilon_{1,c}$ occurs, remaining positive up to the VIS region. These hyperbolic regions unveil the different nature beneath: The first one shows a frequency-confined Drude-Lorentz hyperbolicity in the THz spectral region (see below), while the second hyperbolic region is related to a Drude-Drude mechanism and spans over a wide frequency bandwidth in the IR-VIS range.

In order to assess the quality of a hyperbolic material, four quality parameters can be defined [1,24]. The first one is the hyperbolic operational spectral bandwidth $\Delta\omega$, which is the frequency range over which the hyperbolic behavior is enabled. The second one is the quality factor, $Q(\omega) = -\frac{\epsilon_1(\omega)}{\epsilon_2(\omega)}$, defined as the ratio between the real $\epsilon_1$ and the imaginary value of $\epsilon_2$ of the most conductive axis or layer ($\epsilon_1(\omega) < 0$). The third one, $Q_{\max}$, is the maximum value of $Q(\omega)$ in the $\Delta\omega$ range. Finally, the strength of the dielectric anisotropy (SDA) is defined as the difference $\Delta\varepsilon_1 = \varepsilon_{1,c} - \varepsilon_{1,ab}$ between the real part of the dielectric function along the non-conductive $c$ axis ($\varepsilon_{1,c}$) and that in the $ab$ conductive plane ($\varepsilon_{1,ab}$).

According to the experimental results, PdCoO$_2$ and PdCrO$_2$ show exceptional values for all these parameters in both hyperbolic regions. In particular, we show in Figure 3a and 3b $Q(\omega)$ and $\Delta\varepsilon_1(\omega)$ parameters for both materials. In the first hyperbolic domain (shaded light orange region), the operational spectral bandwidth $\Delta\omega$ of PdCoO$_2$ is about 140 cm$^{-1}$, extending from nearly 580 cm$^{-1}$ to 720 cm$^{-1}$, while the PdCrO$_2$ one is 35 cm$^{-1}$, from 635 cm$^{-1}$ to 670 cm$^{-1}$. Both materials present high $Q$ factors (pink dots in Figure 3c) in this region, comparable with other phononic hyperbolic materials (blue dots in Figure 3c). The dielectric anisotropy $\Delta\varepsilon_1(\omega)$, is clearly visible in Figure 3d (pink dots), reaching extreme values, at least ~30 times higher than other hyperbolic materials.

Concerning the second hyperbolic domain (shaded light blue region), the $\Delta\omega$ of PdCoO$_2$ is about 11200 cm$^{-1}$, extending from nearly 800 cm$^{-1}$ to 12000 cm$^{-1}$, while the $\Delta\omega$ of PdCrO$_2$ is nearly 14300 cm$^{-1}$, from 1300 cm$^{-1}$ to 15600 cm$^{-1}$. These extremely wide hyperbolic operational spectral bandwidths are coupled with huge values of $Q$ factors, which are shown in Figure 3a (shaded light blue region). From Figures 3a and 3b, it is clear that both systems exhibit hyperbolic behavior from the IR to the VIS range, with optimal performance in the mid-IR spectral range. In particular,

PdCoO$_2$ reaches a $Q_{MAX}$ of 23 and a $\Delta\varepsilon_1$ = 135 at ~3000 cm$^{-1}$ while PdCrO$_2$ reaches a $Q_{MAX}$ of 13 and a $\Delta\varepsilon_1$ = 78 at ~4000 cm$^{-1}$.

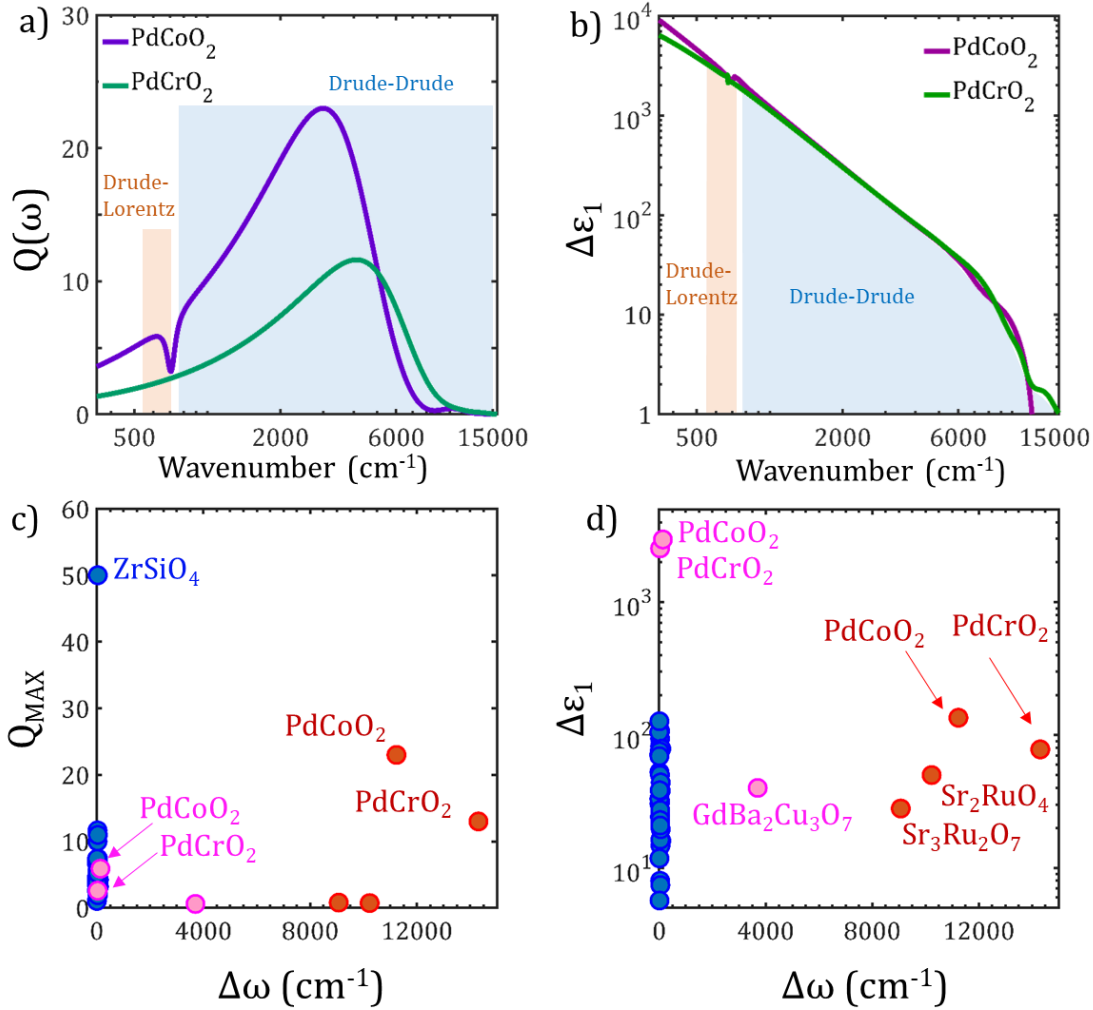

**Figure 3.** $Q$ factor a) and $\Delta\epsilon_1$ b) vs. frequency for PdCoO$_2$ and PdCrO$_2$, extracted from the experimental optical dielectric tensors presented in Figure 2. The shaded regions represent the Drude-Lorentz (light orange) and Drude-Drude (light blue) hyperbolic regions. c) Experimental $Q_{MAX}$ factor vs. the hyperbolic operational spectral bandwidth $\Delta\omega$ for known Lorentz-Lorentz hyperbolic materials (blue dots)[1], Drude-Lorentz hyperbolic materials (pink dots)[25], and Drude-Drude hyperbolic materials (red dots)[25]. d) Strength of the dielectric anisotropy $\Delta\varepsilon_1(\omega)$ vs. $\Delta\omega$ for the same materials.

The above quality factors are far higher than those in other hyperbolic materials[1,24,25]. To emphasize this, in Figure 3c we display a $Q_{MAX}$ vs. $\Delta\omega$ scatter plot, comparing PdCoO$_2$ and PdCrO$_2$ with other hyperbolic systems[1]. The quality parameters plotted in this panel have been extracted from the experimental data found in the literature, and three categories of hyperbolic domain are shown: Lorentz-Lorentz (blue dots)[1], Drude-Lorentz (pink dots)[25], and Drude-Drude (red dots)[25]. Usually, the highest performing hyperbolic materials are the Lorentz-Lorentz ones, which means that their hyperbolic $\Delta\omega$ is very narrow, typically tens of cm$^{-1}$. PdCoO$_2$ and PdCrO$_2$ crystals instead present an extremely large $\Delta\omega$ and impressive $Q_{MAX}$ value (only the ZrSiO$_4$ one is higher, but with much lower $\Delta\omega$ and $\Delta\varepsilon_1$), associated with the very high $\Delta\varepsilon_1$.

All these experimental results indicate that PdCoO$_2$ and PdCrO$_2$ are hyperbolic (Type II) materials with unique quality parameters in the terahertz and infrared spectral region. We now highlight how these properties can be useful for some photonics applications. As discussed in the Introduction, hyperbolic materials support high-$k$ (large spatial frequency) modes due to their hyperbolic dispersion, allowing the propagation of evanescent waves[2–6]. This property plays a key role in hyperlensing and imaging at a nanoscale spatial resolution. Indeed, a hyperlens is not only capable of resolving subwavelength features but also converts subwavelength information (carried by evanescent waves) into propagating waves that can be detected in the far-field regime [6,10]. These materials are also an excellent platform for catalysis[26], and new results on catalysis show the potential role of polaritons to enhance reaction rates [27].

The previously discussed experimental optical characteristics of delafossite crystals also generated very peculiar polariton behavior. Indeed, in the presence of a hyperbolic anisotropic dielectric response, it is possible to excite hyperbolic plasmon (phonon) polaritons (HPPs and HPhPs) [4,5,9]. These excitations merge the characteristics of surface plasmon (phonon) polaritons, which are localized surface waves at an interface[28,29], with the unusual anisotropic dispersion of hyperbolic materials, enabling distinctive control over highly directional light-matter interactions.

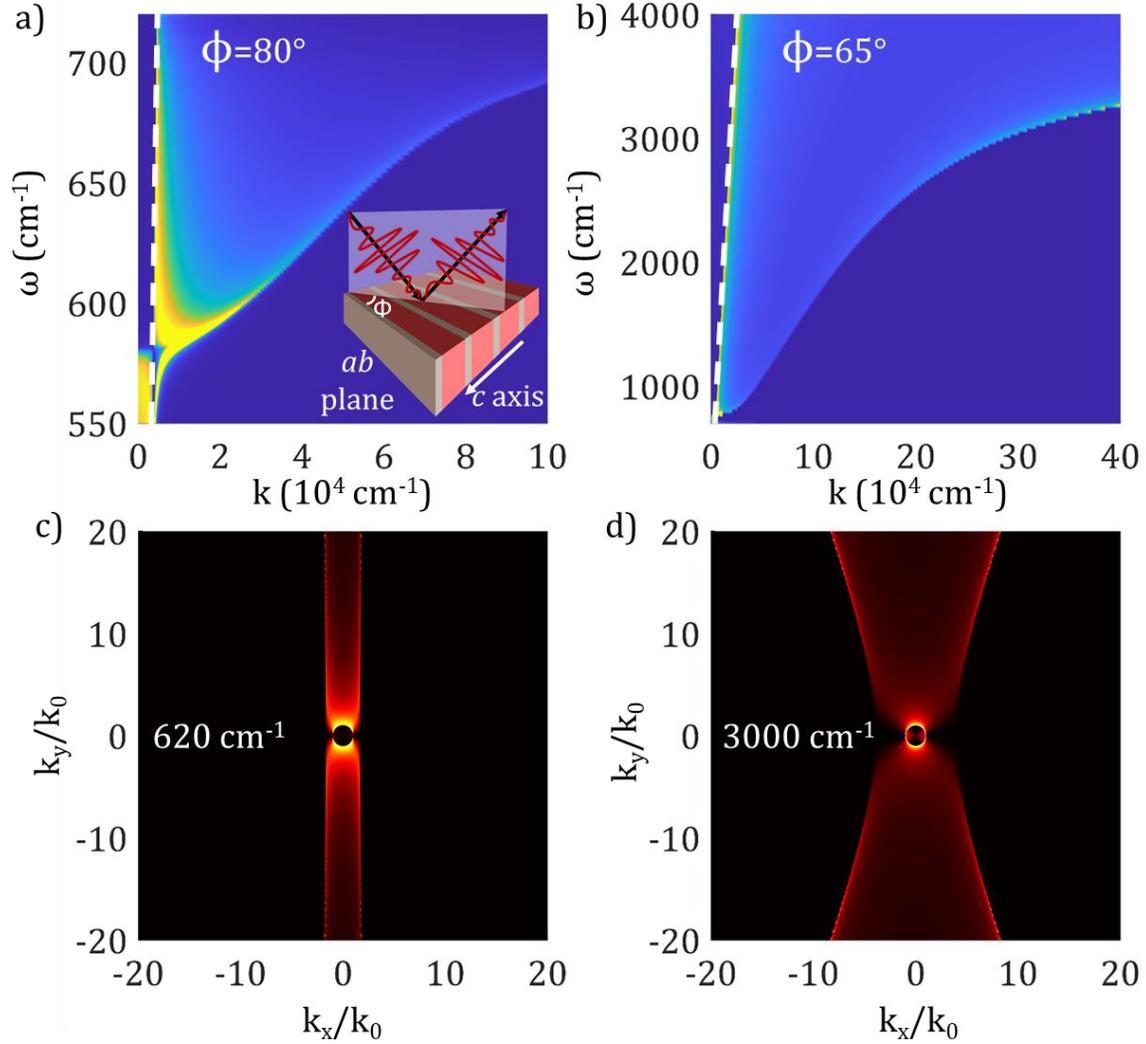

**Figure 4.** Dispersion relation for hyperbolic phonon a) and hyperbolic plasmon b) polaritons in PdCoO$_2$ obtained from the imaginary part of the Fresnel reflection coefficient $Im(r_{pp})$ for $p$ polarized light[4] with different tilt angles: a) $\phi = 80°$ , b) $\phi = 65°$. $\phi$ is the angle formed by the plane of incidence and the conductive $ab$ plane (inset to panel a). The white dashed lines represent the light-line dispersion. In hyperbolic systems, due to their optical anisotropy, the optical response is highly dependent on this angle. c-d) In-plane isofrequency curves at $Q_{MAX}$ frequencies for the Drude-Lorentz hyperbolic (620 cm$^{-1}$), and Drude-Drude (3000 cm$^{-1}$) spectral regions. Momentum values ($k_x$, $k_y$) were normalized to the momentum of the impinging radiation ($k_0$). The circular region in the center of both figures corresponds to an isotropic dispersion in air with radius $k_0$, inside which there is no propagation. In order to enhance the visibility of the curves, we used a custom arbitrary units color scale.

In particular, in hyperbolic systems, the optical response is highly directional, meaning it depends on the orientation of the radiation's incident plane relative to the crystal. To proceed with the calculation of the hyperbolic plasmon and phonon polariton dispersion, we introduce $\phi$, the angle formed between the *ab* plane and the plane of incidence (see the inset in Figure 4a). If we want to excite a HPP resonating at $\omega$, the radiation (so the angle $\phi$) needs to align with the group velocity direction at $\omega$ whose direction is defined by the optical properties of the hyperbolic material[5].

For example, we show in Figure 4 the frequency-momentum dispersions of hyperbolic phonon (Figure 4a) and plasmon (Figure 4b) polaritons, generated on the surface of a PdCoO$_2$ single crystal that contains the *c* axis. Here, the white dashed lines represent the light-line dispersion. The HPP frequency-momentum dispersion is obtained from the poles of the imaginary part of the reflection Fresnel coefficient $Im(r_{pp})$ for p-polarized fields [4,5,30]. The different behaviors presented in Figures 4a and 4b are obtained by changing $\phi$. These angles were selected to match the group velocity directions at the $Q_{MAX}$ frequencies of the Drude-Lorentz hyperbolic regime (Figure 4a, $\phi = 80°$, $\omega_{Qmax}$ = 620 cm$^{-1}$) and Drude-Drude hyperbolic regime (Figure 4b, $\phi = 65°$, $\omega_{Qmax}$ = 3000 cm$^{-1}$).

We finally plot in Figures 4c and 4d the isofrequency maps of PdCoO$_2$ for both hyperbolic regions, which show the $Im(r_{pp})$ for a fixed frequency plotted in $k_x$ and $k_y$ space, where $\hat{x}$ and $\hat{y}$ are the directions perpendicular and parallel to the *c* axis. Momentum values were normalized to the momentum of the impinging radiation $k_0 = \omega/c$. These plots are effectively a cross section of the isofrequency hyperboloid, with at the center a circular region related to the isotropic dispersion in air with radius $k_0$, inside which there is no propagation. Figure 4c shows the almost flat dispersion obtained at a $Q_{MAX}$ frequency of 620 cm$^{-1}$ for the Drude-Lorentz hyperbolic regime, with group velocity direction almost horizontal. Figure 4d instead displays the Drude-Drude hyperbolic regime at 3000 cm$^{-1}$.

These calculations underline again how these crystals have impressive performance suitable for wide-range infrared plasmonic applications. By controlling $\phi$, it is possible to optimize the coupling between the incident light and the hyperbolic modes, enabling highly controlled and directional propagation of these surface waves.

## 3. Conclusion

In this paper, we have measured the dielectric tensor of delafossite-metals $PdCoO_2$ and $PdCrO_2$ single crystals over a broad spectral range from the terahertz to visible. Both delafossite crystals exhibit a strong optical metallic response in their *ab* plane, showing a plasma edge in the near-infrared. In contrast, along the *c* axis they show a strong phonon absorption, which is superimposed on weakly metallic behavior. The interplay between these two distinct optical responses gives rise to a dual hyperbolic regime whose primary metrics - the operational spectral bandwidth $\Delta\omega$, the quality factor $Q$, and the strength of the dielectric anisotropy $\Delta\epsilon$ – achieve record values with respect to any known hyperbolic systems. Using the experimental dielectric functions, we calculated the frequency-momentum dispersions of surface hyperbolic phonon and plasmon polaritons mapping the isofrequency curves in both hyperbolic regimes. In the hyperbolic materials landscape, these results clearly highlight the exceptional performances of these delafossite layered metals, which enable highly controlled and directional propagation of hyperbolic modes, offering clear benefits for photonic and plasmonic applications.

## 4. Methods

### 4.1 Materials

Crystal growth methods were briefly summarized in Section 2 and are described in detail in Supporting Information. In essence, metathesis reactions were employed to generate precursor $PdCoO_2$ and $PdCrO_2$ powders, followed by chemical vapor transport (CVT) single crystal growth with a $PdCl_2$ transport agent. As described in detail in Supporting Information (see Figure S1), structural and chemical characterization were performed by single-crystal X-ray diffraction, powder X-ray diffraction on ground crystals, and energy dispersive X-ray spectroscopy. These techniques confirm single crystallinity, phase purity, accepted lattice parameters, and nominal stoichiometry. Deeper characterization has been performed for these $PdCoO_2$ crystals previously, in particular establishing ~50 ppm metal-basis total impurity concentrations, ultrapure Pd planes, and record residual resistivity ratios of >440[12].

## 4.2 Optical Measurements

Optical measurements in the THz-VIS range were carried out with a Bruker Vertex 70v spectrometer coupled with a Hyperion 1000 microscope, scanning from 190 cm$^{-1}$ up to 15000 cm$^{-1}$ with a resolution of 4 cm$^{-1}$. A set of THz-IR-VIS polarizers were used to control the electric field direction with respect to the crystal axis. The optical data were analyzed with the RefFIT software[20], in order to extract the real and imaginary parts of the dielectric function.

## 4.3 Numerical models

The HPP's frequency-momentum dispersion and the planar isofrequency curves have been obtained by numerical evaluation of the poles of the imaginary part of the reflection Fresnel coefficient for p-polarized fields, $Im(r_{pp})$[4,5], at the air/crystal interface for both propagating and evanescent wave excitation. Calculations were performed using a transfer matrix method (TMM) for anisotropic, dissipative layered structures [30]. A PdCoO$_2$ single crystal is considered and its dielectric tensor is oriented as follows:

$$\begin{pmatrix} \tilde{\epsilon}_{ab} & 0 & 0 \\ 0 & \tilde{\epsilon}_{c} & 0 \\ 0 & 0 & \tilde{\epsilon}_{ab} \end{pmatrix} \quad (2)$$

Where $\tilde{\epsilon}_{ab} = \epsilon_{1,ab} + i\epsilon_{2,ab}$, $(\tilde{\epsilon}_{c} = \epsilon_{1,c} + i\epsilon_{2,c})$ are the experimental complex dielectric functions along the *ab* plane (*c* axis), represented in Figures 2c (Figure 2a). Thus, the *c* axis is parallel to the *y* axis while the unit vector normal to the air/crystal interface is parallel to the *z* axis. We define $\phi$ as the angle between the *c* axis and the unit vector normal to the incidence plane. The isofrequency curves are calculated by rotating the plane of incidence along the *z* axis and evaluating the corresponding Fresnel reflection coefficient as a function of the in-plane momentum components ($k_x$, $k_y$) at fixed frequency.

## Contributions

All authors contributed extensively to this work. S.M. and S.L. designed the experiment. S.L. and S.M. supervised the work and the research program. Y.T., P.J., Y.Z., F.T., and C.L. grew and characterized the $PdCoO_2$ and $PdCrO_2$ single crystals. V.S. prepared the sample surfaces. S.M., A.D., L.M., and C.P. performed the infrared spectroscopy and analysed the data. M. C., M. C. L. and E.D.R. developed and performed the numerical simulations. S.M. L.M., and S.L. prepared the original draft. All authors reviewed and edited the paper. All authors have read and agreed to the published version of the manuscript.

**Acknowledgment**

This publication was supported by the European Union under the Italian National Recovery and Resilience Plan (NRRP) of Next Generation EU partnership PE0000023-NQSTI, and MUR PRIN project PHOtonics Terahertz devices based on tOpological materials (PHOTO) 2020RPEPNH. All work at the University of Minnesota (crystal growth and characterization) was supported by the US Department of Energy through the University of Minnesota (UMN) Center for Quantum Materials under DE-SC0016371. Parts of this work were conducted in the UMN Characterization Facility, which is partially supported by the US National Science Foundation through the MRSEC program under DMR-2011401. We acknowledge support through the PRIN 2022 MUR Project No. 20223T577Z.

**References**

[1] K. Korzeb, M. Gajc, D. A. Pawlak, *Opt. Express, OE* **2015**, *23*, 25406.
[2] A. Poddubny, I. Iorsh, P. Belov, Y. Kivshar, *Nature Photon* **2013**, *7*, 948.
[3] L. Ferrari, C. Wu, D. Lepage, X. Zhang, Z. Liu, *Progress in Quantum Electronics* **2015**, *40*, 1.
[4] S. Dai, Z. Fei, Q. Ma, A. S. Rodin, M. Wagner, A. S. McLeod, M. K. Liu, W. Gannett, W. Regan, K. Watanabe, T. Taniguchi, M. Thiemens, G. Dominguez, A. H. C. Neto, A. Zettl, F. Keilmann, P. Jarillo-Herrero, M. M. Fogler, D. N. Basov, *Science* **2014**, *343*, 1125.
[5] Y. Shao, A. J. Sternbach, B. S. Y. Kim, A. A. Rikhter, X. Xu, U. De Giovannini, R. Jing, S. H. Chae, Z. Sun, S. H. Lee, Y. Zhu, Z. Mao, J. C. Hone, R. Queiroz, A. J. Millis, P. J. Schuck, A. Rubio, M. M. Fogler, D. N. Basov, *Science Advances* **2022**, *8*, eadd6169.
[6] Y. Gelkop, F. Di Mei, S. Frishman, Y. Garcia, L. Falsi, G. Perepelitsa, C. Conti, E. DelRe, A. J. Agranat, *Nat Commun* **2021**, *12*, 7241.
[7] S. Abedini Dereshgi, M. C. Larciprete, M. Centini, A. A. Murthy, K. Tang, J. Wu, V. P. Dravid, K. Aydin, *ACS Appl. Mater. Interfaces* **2021**, *13*, 48981.

[8]  M. C. Larciprete, S. A. Dereshgi, M. Centini, K. Aydin, *Opt. Express, OE* **2022**, *30*, 12788.
[9]  G. Venturi, A. Mancini, N. Melchioni, S. Chiodini, A. Ambrosio, *Nat Commun* **2024**, *15*, 9727.
[10] O. Takayama, A. V. Lavrinenko, *J. Opt. Soc. Am. B, JOSAB* **2019**, *36*, F38.
[11] Y. Zhang, *Phys. Rev. Mater.* **2022**, *6*.
[12] Y. Zhang, F. Tutt, G. N. Evans, P. Sharma, G. Haugstad, B. Kaiser, J. Ramberger, S. Bayliff, Y. Tao, M. Manno, J. Garcia-Barriocanal, V. Chaturvedi, R. M. Fernandes, T. Birol, W. E. Seyfried, C. Leighton, *Nat Commun* **2024**, *15*, 1399.
[13] R. D. Shannon, D. B. Rogers, C. T. Prewitt, *Inorg. Chem.* **1971**, *10*, 713.
[14] J.-P. Doumerc, A. Wichainchai, A. Ammar, M. Pouchard, P. Hagenmuller, *Materials Research Bulletin* **1986**, *21*, 745.
[15] C. C. Homes, *Phys. Rev. B* **2019**, *99*.
[16] H. Takatsu, Y. Maeno, *Journal of Crystal Growth* **2010**, *312*, 3461.
[17] H. Takatsu, S. Yonezawa, C. Michioka, K. Yoshimura, Y. Maeno, *J. Phys.: Conf. Ser.* **2010**, *200*, 012198.
[18] MDI JADE 9.1, *MDI JADE 9.1*, **2019**.
[19] J.-P. Doumerc, A. Ammar, A. Wichainchai, M. Pouchard, P. Hagenmuller, *Journal of Physics and Chemistry of Solids* **1987**, *48*, 37.
[20] A. B. Kuzmenko, *Review of Scientific Instruments* **2005**, *76*, 083108.
[21] A. P. Mackenzie, *Rep. Prog. Phys.* **2017**, *80*, 032501.
[22] H. Takatsu, S. Yonezawa, S. Mouri, S. Nakatsuji, K. Tanaka, Y. Maeno, *J. Phys. Soc. Jpn.* **2007**, *76*, 104701.
[23] J. M. Ok, M. Brahlek, W. S. Choi, K. M. Roccapriore, M. F. Chisholm, S. Kim, C. Sohn, E. Skoropata, S. Yoon, J. S. Kim, H. N. Lee, *APL Materials* **2020**, *8*, 051104.
[24] G. Jia, J. Luo, H. Wang, Q. Ma, Q. Liu, H. Dai, R. Asgari, *Nanoscale* **2022**, *14*, 17096.
[25] J. Sun, N. M. Litchinitser, J. Zhou, *ACS Photonics* **2014**, *1*, 293.
[26] G. Li, S. Khim, C. S. Chang, C. Fu, N. Nandi, F. Li, Q. Yang, G. R. Blake, S. Parkin, G. Auffermann, Y. Sun, D. A. Muller, A. P. Mackenzie, C. Felser, *ACS Energy Lett.* **2019**, *4*, 2185.
[27] B. Xiang, W. Xiong, *Chem. Rev.* **2024**, *124*, 2512.
[28] S. Macis, L. Tomarchio, S. Tofani, F. Piccirilli, M. Zacchigna, V. Aglieri, A. Toma, G. Rimal, S. Oh, S. Lupi, *Commun Phys* **2022**, *5*, 1.
[29] S. Macis, A. D'Arco, L. Mosesso, M. C. Paolozzi, S. Tofani, L. Tomarchio, P. P. Tummala, S. Ghomi, V. Stopponi, E. Bonaventura, C. Massetti, D. Codegoni, A. Serafini, P. Targa, M. Zacchigna, A. Lamperti, C. Martella, A. Molle, S. Lupi, *Advanced Materials* **2024**, *36*, 2400554.
[30] N. C. Passler, A. Paarmann, *J. Opt. Soc. Am. B, JOSAB* **2017**, *34*, 2128.